\documentclass[a4paper]{article}

\usepackage{INTERSPEECH2022}

\usepackage{multirow}

\title{Exploring Timbre Disentanglement in Non-Autoregressive Cross-Lingual Text-to-Speech}
\name{Haoyue Zhan, Xinyuan Yu, Haitong Zhang, Yang Zhang, Yue Lin}
\address{
  NetEase Games AI Lab, Guangzhou, China
  }
\email{\{zhanhaoyue, yuxinyuan02, zhanghaitong01, zhangyang09, gzlinyue\}@corp.netease.com}

\begin{document}

\maketitle

\begin{abstract}
In this paper, we study the disentanglement of speaker and language representations in non-autoregressive cross-lingual TTS models from various aspects.
We propose a phoneme length regulator that solves the length mismatch problem between IPA input sequence and monolingual alignment results.
Using the phoneme length regulator, we present a FastPitch-based cross-lingual model with IPA symbols as input representations.
Our experiments show that language-independent input representations (e.g. IPA symbols), an increasing number of training speakers, and explicit modeling of speech variance information all encourage non-autoregressive cross-lingual TTS model to disentangle speaker and language representations.
The subjective evaluation shows that our proposed model can achieve decent naturalness and speaker similarity in cross-language voice cloning. 
\end{abstract}

\section{Introduction}
\label{sec:intro}

In the past years, end-to-end Text-to-Speech (TTS) synthesis systems have gained great success in generating natural monolingual speech \cite{shen2018natural, ren2019fastspeech, ren2021fastspeech, FastPitch}. However, for a deployed TTS system, it is very common to synthesize mixed language utterances. A recent review \cite{do21_interspeech} further shows that cross-lingual TTS systems can help boost the quality of synthesized speech for low-resource language. 
Nevertheless, state-of-the-art TTS models still have a gap in generating natural cross-lingual utterances, especially when only monolingual training data are available.


One of the difficulties in building a cross-lingual TTS model with monolingual data lies in that the speaker representations and language representations are often entangled with each other. On one hand, various adversarial methods are adopted to ease this problem. \cite{zhang2019learning} employs domain adversarial training to disentangle the text and speaker representations. In \cite{xin20_interspeech}, the authors use domain adaptation and perceptual similarity regression to find similar cross-lingual speaker pairs to build cross-lingual TTS models. 
\cite{shang21_interspeech} introduces domain adversarial training into the non-autoregressive acoustic model, and builds a multi-speaker multi-style multilingual TTS system. On the other hand, some researchers have studied the implication of input representations for cross-lingual TTS systems. \cite{liu20m_interspeech} builds a shared phoneme set for three different languages. \cite{chen19f_interspeech} proposes a phonetic transformation network to learn target symbol distribution with the help of Automatic Speech Recognition (ASR) systems. In \cite{9053094, zhao20e_interspeech}, language-independent Phonetic PosteriorGram (PPG) features of ASR models are used as input for cross-lingual TTS models. \cite{bansal20_interspeech} further proposes a mixed-lingual grapheme-to-phoneme (G2P) frontend to improve the pronunciation of mixed-lingual sentences in cross-lingual TTS systems. 

Besides language-dependent representation, some researches focus on using language-independent representation as input representations. \cite{chen2019cross} uses International Phonetic Alphabet (IPA) as text input representations and adopts a ResCNN-based speaker encoder to encode speaker representations. \cite{fu20b_interspeech} combines IPA with dynamic soft windowing mechanism and language-dependent style token to improve intelligibility, naturalness, and speaker similarity of code-switching speech. In \cite{staib20_interspeech, maniati21_interspeech}, the authors propose to transform IPA to phonological features to build cross-lingual TTS models. Staib \textit{et al.} \cite{staib20_interspeech} even extend the model to an unseen language. These studies show that language-independent representations can simplify the training procedure of cross-lingual TTS models and help disentangle speaker and language representations.

Recently, non-autoregressive TTS models \cite{FastPitch, ren2021fastspeech} have achieved state-of-the-art performance for monolingual TTS in terms of speech intelligibility and naturalness. Nonetheless, most of the aforementioned works adopt autoregressive TTS models as their backbone framework, and few works exploit non-autoregressive architecture in cross-lingual TTS models. 
One important reason is that typical non-autoregressive TTS models contain separate duration modules which depend on external aligners \cite{ren2021fastspeech} to provide ground truth labels for training. This impedes the use of language-independent representation (e.g. IPA symbols) as model input. To fill this research gap, in this paper, we seek to adopt non-autoregressive architecture to build cross-lingual TTS models. We propose a phoneme length regulator to integrate IPA input representations into non-autoregressive TTS models, and study how different input representations contribute to speaker disentanglement from language. 
We scale up the numbers of training speakers to investigate its impact on timbre disentanglement for non-autoregressive cross-lingual TTS models. Throughout extensive experiments, we find that a FastPitch-based \cite{FastPitch} cross-lingual model with IPA symbols as input representations achieves the best speech naturalness and speaker similarity. We further verify the effectiveness of each component of the model by ablation studies.

 \begin{figure}[htb]
   \centering
   \includegraphics[width=\linewidth]{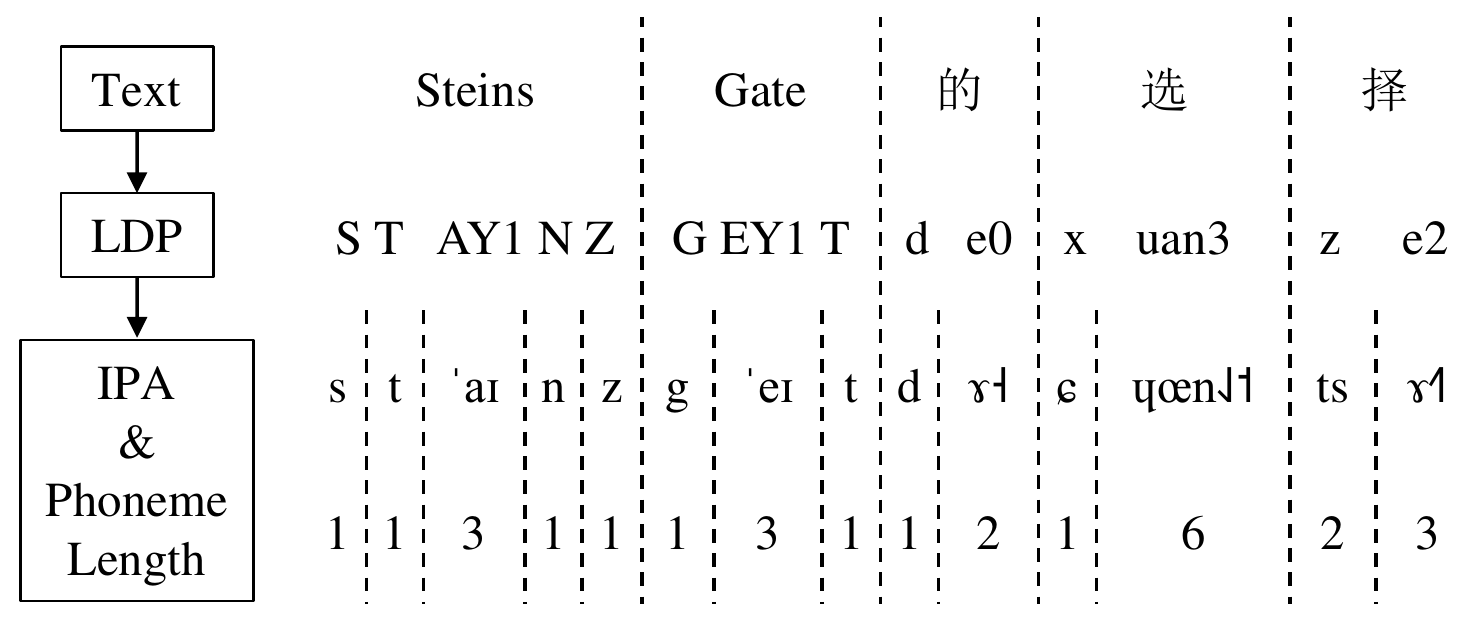}
   \caption{The process of converting language-dependent phoneme sequence to IPA sequence.}
   \label{fig:G2P}
 \end{figure}

Our contributions are as follows. 
(1) We propose a phoneme length regulator to build non-autoregressive cross-lingual TTS models with IPA input.
(2) We evaluate the effect of different numbers of training speakers on timbre disentanglement for non-autoregressive cross-lingual TTS models. 
(3) We study the impact of adversarial training and variance adaptors on naturalness and speaker similarity for non-autoregressive cross-lingual TTS models.


\section{Proposed Approach}
\label{sec:proposed}

\subsection{Mapping language-dependent phoneme to IPA symbols}
\label{sec:map_ipa}
For any given text sequence, we first convert each word or syllable to its language dependent phoneme. Then, we use a custom dictionary $\footnote{https://github.com/open-dsl-dict/ipa-dict-dsl}$ to map each language-dependent phoneme to IPA symbols. 
Since IPA symbols are fine-grained phonetic notations, one language-dependent phoneme (LDP) can usually be decomposed into one or more IPA symbols. We refer to the number of IPA symbols of the corresponding LDP as the phoneme length. 

Figure \ref{fig:G2P} illustrates the above process with a sentence containing both Mandarin and English. In the first two rows of Figure \ref{fig:G2P}, each word is converted to either ARPABET for English or Pinyin for Mandarin. In the last two rows, we map each ARPABET or Pinyin symbol to its corresponding IPA symbols and phoneme length. 

\subsection{Phoneme length regulator}
\label{sec:PLR}

Different from autoregressive TTS models, most non-autoregressive TTS models \cite{FastPitch, ren2021fastspeech} rely on external force aligners to provide phoneme duration information during training. 
However, IPA sequences have different lengths from the LDP sequences and cannot use the monolingual alignment results.
To solve this length mismatch problem, we propose a phoneme length regulator to convert IPA embeddings back to LDP embeddings, which effectively bridges the IPA input sequences and the monolingual alignment results. 
Following the process described in Section \ref{sec:map_ipa}, given the IPA sequence $\mathbf{X}=\{X_j\}_{j=1}^{T_X}$ and the corresponding phoneme length sequence $\mathbf{L}=\{L_i\}_{i=1}^{T_L}$, the phoneme length regulator outputs the aggregated embedding sequence $\mathbf{Y}$ by adding the IPA embeddings corresponding to the same language-dependent phoneme based on the phoneme length sequence as:
\begin{equation}
\left\{
    \begin{array}{ll}
         c_\tau = 0, & \tau=0  \\
         c_\tau = \sum_{k=1}^{\tau}L_k, & 0 < \tau \leq T_L \\
    \end{array}
\right.
\end{equation}
\begin{equation}
    Y_i = \sum_{k=c_{i-1}+1}^{c_i} X_k, \quad for \; 1 \leq i \leq T_L
\end{equation}
Where $c_\tau$ is the cumulative sum of the phoneme length sequence and $\mathbf{Y}=\{Y_i\}_{i=1}^{T_L}$ is the aggregated embedding sequence.
One may see an aggregated embedding as a language-dependent embedding, because it's aggregated from the IPA embeddings corresponding to the same language-dependent phoneme. 
Now that the aggregated sequence has a one-to-one corresponding relationship to the original LDP sequence, one can use the duration information provided by a monolingual aligner to train non-autoregressive TTS models.

 \begin{figure}[htb]
   \centering
   \includegraphics[width=\linewidth]{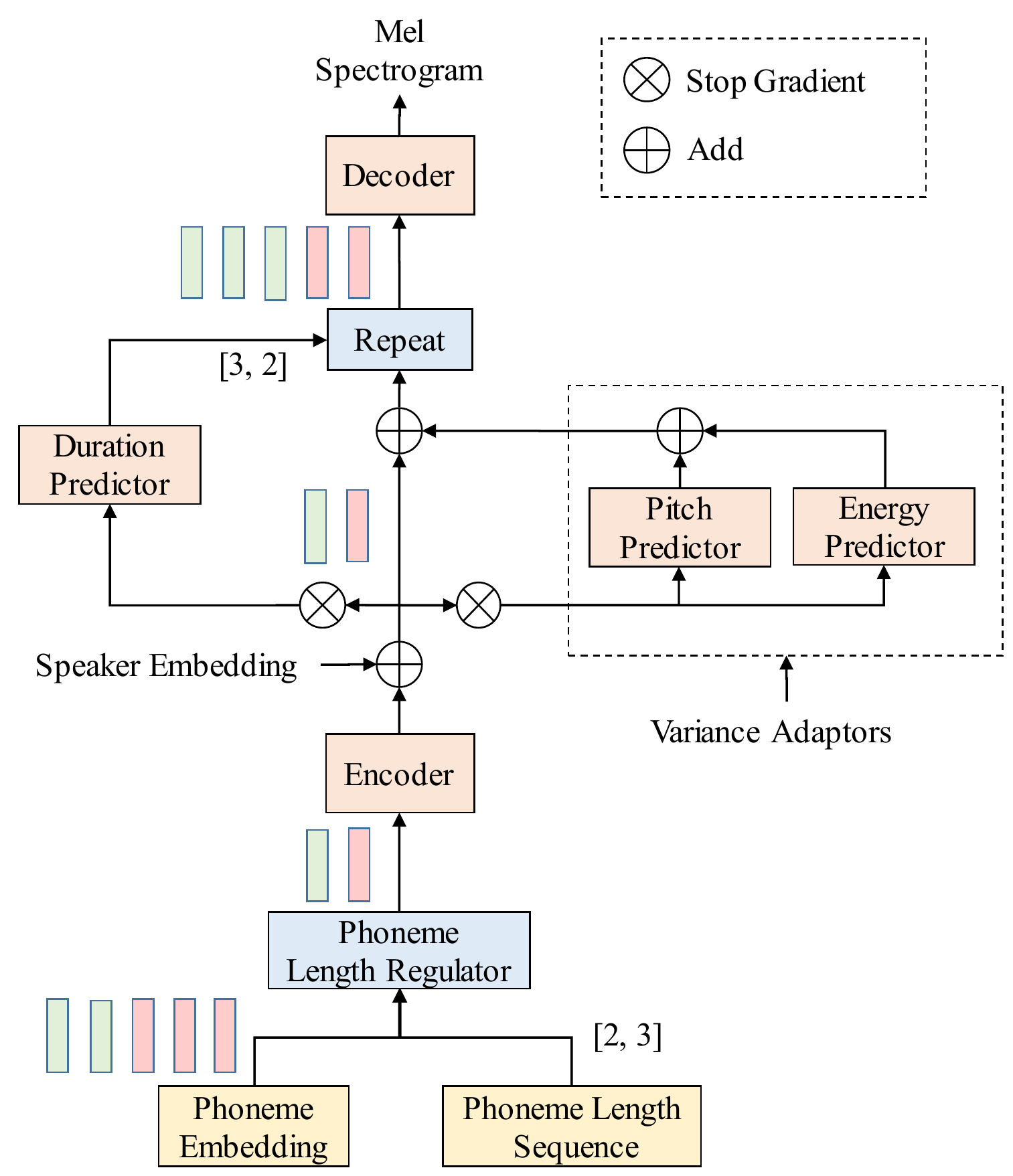}
   \caption{Overview of the proposed model.}
   \label{fig:model}
 \end{figure}

\subsection{Core acoustic model}
\label{sec:propose_model}
The proposed phoneme length regulator in Section \ref{sec:PLR} can be easily applied to any existing non-autoregressive models to enable IPA input. In this section, we propose a FastPitch-based \cite{FastPitch} cross-lingual model with IPA symbols as input.
We show in Section \ref{sec:exper} that, when trained on a multi-speaker cross-lingual dataset, the proposed model achieves the best MOS score among other strong baselines.



As shown in Figure \ref{fig:model}, the backbone of the proposed model is inherited from FastPitch \cite{FastPitch}. 
The main architecture of the proposed model consists of an encoder and a decoder. Both the encoder and the decoder have 4 layers of Feed-Forward Transformer (FFT) blocks.
However, unlike \cite{FastPitch} which adds speaker embedding to the encoder input, we add trainable speaker embedding to the encoder output.

In addition, the original FastPitch model only use a pitch predictor to improve the quality of generated speech. To increase the stability of our proposed model, we add an extra energy predictor as in \cite{ren2021fastspeech}.
To ensure the training of these variance adaptors will not negatively affect the training of encoder \cite{glow_tts,raitio20_interspeech,9414720}, we add a stop-gradient operation to the input of all variance adaptors as shown in the middle part of Figure \ref{fig:model}. 

Frame-aligned energy is calculated in the same way as \cite{ren2021fastspeech}. We then average the energy frames that correspond to the same language-dependent phoneme based on extracted duration, and quantize energy in the same way as \cite{ren2021fastspeech}.
Meanwhile, we obtain ground truth pitch value through WORLD \cite{morise2016world} and average the pitch value using the same method as energy. Instead of quantizing the pitch value, we follow \cite{FastPitch} and use a 1-D convolution layer to convert pitch value into pitch embedding.
The variance adaptors and the main model are all optimized with mean-squared error (MSE) loss similar to \cite{FastPitch, ren2021fastspeech}.



\section{Experiment and Results}
\label{sec:exper}

\subsection{Dataset setup}
\label{sec:datasetup}

\begin{table}[htb]
\centering
\caption{Details of training datasets.}
\begin{tabular}{cccc}
\toprule
Dataset & Speaker & Source & Duration (Used) \\
\midrule
\multirow{2}{*}{$d1$} & CN male & Internal & 5 hours\\ 
                  & EN female       & LJSpeech \cite{ljspeech17}      & 5 hours     \\ \hline
\multirow{2}{*}{$d2$} & CN female & Databaker \cite{BIAOBEI}  & 1 hour\\ 
                  & EN male       & cmu\_bdl \cite{kominek2004cmu}      & 1 hour      \\ \hline
\multirow{4}{*}{$d3$} &  CN male & Internal &  1 hour\\
                  & CN female       & Internal      &  1 hour \\ 
                  &  EN male & cmu\_rms \cite{kominek2004cmu} &  1 hour \\
                  &  EN female & cmu\_clb \cite{kominek2004cmu} &  1 hour  \\
\bottomrule      
\end{tabular}
\label{tab:table1}
\end{table}

We conduct all the experiments in two languages, Mandarin (CN) and English (EN). We use the LJSpeech \cite{ljspeech17} dataset (EN), the Databaker \cite{BIAOBEI} dataset (CN), the CMU arctic \cite{kominek2004cmu} dataset (EN) and our proprietary dataset (CN). 
We take subsets from the aforementioned datasets, and combine them into three cross-lingual datasets, namely $d1$, $d2$, and $d3$. All models are trained on one or more of the three datasets. 
Details of the three datasets are shown in Table \ref{tab:table1}. 
We carefully choose the gender and language of the speakers so that the language and gender of the datasets are as balanced as possible \cite{do21_interspeech}. Note that the speakers in $d1$ have the most training data. Thus, we consider the two speakers as our target voices and denote them as $d1$-CN-M and $d1$-EN-F, respectively.

\subsection{Experimental setup}
\label{sec:exp_setup}


All speech data are resampled at 16kHz. The audio features are represented as a sequence of 80-dim log-mel spectrogram frames, computed from 40ms windows shifted by 10ms.
The hidden size of FFT blocks of our proposed model is set to 256. Each feed-forward layer of the FFT blocks consists of 2 1-D convolution layers with kernel size 9 and 1024 intermediate channels. The variance adaptors, including the duration predictor, follow that of \cite{ren2021fastspeech}. In all experiments, the proposed model is trained with a batch size of 32 using Adam optimizer for 200k steps. The initial learning rate is set to 0.001, and we apply the same learning rate schedule as \cite{zhan21_interspeech}. 

We compare the \textbf{Proposed} model with the following baseline models. (1) A \textbf{Tacotron-based} cross-lingual model re-implemented from \cite{zhan21_interspeech}. (2) A \textbf{FastSpeech-LDP} cross-lingual model which is a FastSpeech \cite{ren2019fastspeech} based model with language-dependent phoneme as input and adversarial training method. This model is implemented based on the M3 w/o FSE model described in \cite{shang21_interspeech}. (3) A \textbf{FastSpeech-IPA} cross-lingual model, which adopts the same phoneme length regulator and IPA input as the proposed model, but has no variance adaptors.
Notably, for a fair comparison with \textbf{FastSpeech-LDP}, we also apply adversarial training method to \textbf{FastSpeech-IPA}.
All baseline models are trained with the same hyper-parameter settings as the \textbf{Proposed} model.
All generated mel-spectrograms are converted to speech using a universal and fine-tuned HiFi-GAN vocoder \cite{hifigan}.

\subsection{Evaluation}

To evaluate the trained models, we conduct Mean Opinion Score (MOS) tests on speech naturalness and speaker similarity. We select 50 Mandarin (CN) utterances, 50 English (EN) utterances, and 50 mixed-lingual (CN-EN) utterances, and generate speech samples with the selected utterances using both speaker $d1$-CN-M and speaker $d1$-EN-F in Section \ref{sec:datasetup}.
All generated speech utterances are rated by 15 human raters on a scale from 1 to 5 with 0.5 point increments. All raters are native Mandarin speakers with basic English skills.

For speaker similarity, a reference utterance of the same speaker is selected, and raters are instructed to judge whether the given synthesized utterance and reference utterance are spoken by the same person or not. 5 points indicate that the voices of the utterances are definitely the same as the reference utterance whereas 1 point indicates that the voices of the utterances are definitely not the same as the reference utterance. Note that it's difficult to ask raters to ignore the content of generated speech, especially when raters are listening to a non-native language. Thus, bad intelligibility and naturalness of speech might result in a lower MOS score.
Our synthesized speech samples can be found on this website $\footnote{https://hyzhan.github.io/NAC-TTS/}$.


\begin{table}[htb]
\centering
\caption{Naturalness and speaker similarity of speaker $d1$-CN-M in different training datasets.}
\begin{tabular}{p{12mm}p{12mm}p{12mm}p{12mm}p{12mm}p{12mm}}
\toprule
\multirow{2}{*}{MOS} & \multirow{2}{*}{Dataset} &  \multicolumn{3}{c}{Utterances Type} \\ \cline{3-5}
& & \multicolumn{1}{c}{CN} & \multicolumn{1}{c}{CN-EN} & \multicolumn{1}{c}{EN} \\ \midrule

\multirow{3}{*}{Naturalness} & $d1$ & 4.46$\pm$0.15    & 3.61$\pm$0.30 & 3.37$\pm$0.34 \\
                            & $d1$+$d2$ & 4.49$\pm$0.15    & 3.88$\pm$0.21 & 3.64$\pm$0.27 \\
                            & $d1$+$d2$+$d3$ 
  & \textbf{4.51$\pm$0.15}   & \textbf{4.00$\pm$0.19}  & \textbf{3.84$\pm$0.21}\\  \hline
\multirow{3}{*}{Similarity} & $d1$ & \textbf{4.07$\pm$0.36}    & 3.20$\pm$0.42 & 1.76$\pm$0.35\\
                            & $d1$+$d2$ & 4.05$\pm$0.37    & 3.52$\pm$0.36 & 2.97$\pm$0.36\\
                            & $d1$+$d2$+$d3$ 
  & 4.05$\pm$0.36   & \textbf{3.61$\pm$0.35}  & \textbf{3.21$\pm$0.36}
                   \\  \bottomrule
\end{tabular}
\label{tab:table2}
\end{table}

\subsection{Effect of numbers of training speakers}
\label{sec:exp1}
We first investigate the effect of different numbers of training speakers on speaker and language disentanglement. We train our proposed model in Section \ref{sec:propose_model} on different datasets with different numbers of speakers, and evaluate its performance on speaker $d1$-CN-M. The result is presented in Table \ref{tab:table2}.

As shown in Table \ref{tab:table2}, model trained on $d1$ (i.e. only contains the target voices) has the lowest MOS score for foreign language (EN). Whereas the MOS score for native language (CN) reaches the marginal level. This implies that datasets with only one speaker for a language have little help in improving the target voice's ability to speak foreign languages.
We infer that although IPA symbols are shared by all languages, there are still some IPA symbols that are unique to certain languages, which entangles speaker and language representations of the model when the dataset only consists of one speaker for a language.

At the same time, increasing the number of training speakers constantly improves the model's performance of cross-language voice cloning.
In addition, the standard deviation of the naturalness scores also becomes more stable as the number of training speakers increases. 
These results show that the diversity of different speakers not only help the model learn to disentangle language and speaker representations, but also stabilize the generated results of non-autoregressive cross-lingual TTS models. 
Given the abovementioned findings, we train all the models on dataset $d1+d2+d3$ for our experiments in the rest of the paper.

\begin{table*}[htb]
\centering
\caption{Naturalness and speaker similarity of different kinds of utterances for all models.}
\begin{tabular}{cccccccccc}
\toprule
\multirow{2}{*}{Target Speaker} & \multirow{2}{*}{Model} & \multicolumn{2}{c}{CN utterances}  & \multicolumn{2}{c}{CN-EN utterances}
& \multicolumn{2}{c}{EN utterances} \\ \cline{3-8}  
& & Naturalness   & Similarity & Naturalness   & Similarity & Naturalness   & Similarity
\\
\midrule
\multirow{3}{*}{$d1$-CN-M} 
& Tacotron-based \cite{zhan21_interspeech} 
& 2.74$\pm$0.33  & 3.10$\pm$0.42  & 2.57$\pm$0.36  & 2.90$\pm$0.40 & 2.51$\pm$0.35  & 2.53$\pm$0.37   \\
& FastSpeech-LDP \cite{shang21_interspeech}        
& 4.41$\pm$0.19  & 4.16$\pm$0.33  & 3.63$\pm$0.29  & 3.34$\pm$0.39  & 3.51$\pm$0.34  & 2.23$\pm$0.42\\
& FastSpeech-IPA
& 4.44$\pm$0.18  & \textbf{4.17$\pm$0.34}  & 3.75$\pm$0.28  & 3.57$\pm$0.38 & 3.51$\pm$0.32  & 2.20$\pm$0.41   \\
& Proposed         
& \textbf{4.46$\pm$0.18}  & 4.15$\pm$0.34 & \textbf{4.02$\pm$0.24}  & \textbf{3.69$\pm$0.36} & \textbf{3.97$\pm$0.23}  & \textbf{3.40$\pm$0.37} \\
\hline
\multirow{3}{*}{$d1$-EN-F} 
& Tacotron-based \cite{zhan21_interspeech} 
& 2.77$\pm$0.35  & 2.95$\pm$0.43  & 2.85$\pm$0.37  & 3.06$\pm$0.44  & 2.93$\pm$0.39  & 3.11$\pm$0.46 \\
& FastSpeech-LDP \cite{shang21_interspeech}         
& 3.11$\pm$0.33  & 2.13$\pm$0.42  & 3.53$\pm$0.30  & 3.22$\pm$0.45  & 3.88$\pm$0.30  & 3.66$\pm$0.43  \\
& FastSpeech-IPA
& 3.28$\pm$0.33  & 2.72$\pm$0.44  & 3.67$\pm$0.29  & 3.38$\pm$0.42 & 3.92$\pm$0.28  & 3.72$\pm$0.42   \\
& Proposed       
& \textbf{3.92$\pm$0.22}  & \textbf{3.26$\pm$0.42} & \textbf{3.95$\pm$0.20}  & \textbf{3.54$\pm$0.41} & \textbf{4.15$\pm$0.23}  & \textbf{3.80$\pm$0.41} \\ 
\hline

\end{tabular}
\label{tab:table4}
\end{table*}

\subsection{Comparing with baseline models}


Table \ref{tab:table4} presents the result of speech naturalness and speaker similarity MOS tests for the proposed model and all the baseline models in Section \ref{sec:exp_setup}. Generally speaking, non-autoregressive models all outperforms the Tacotron-based baseline model. This result shows the effectiveness of non-autoregressive architecture in cross-lingual TTS.

It can be observed from Table \ref{tab:table4} that FastSpeech-IPA has an overall better speaker similarity score than FastSpeech-LDP. This suggests that using language-independent input representations can better disentangle language and speaker information for non-autoregressive TTS models. In addition, the speech naturalness score of FastSpeech-IPA is no worse or even better than FastSpeech-LDP. This indicates that the proposed phoneme length regulator in Section \ref{sec:PLR} effectively maps IPA embeddings back to LDP embeddings and the model learns the pronunciation of different languages.

Nonetheless, FastSpeech-IPA and FastSpeech-LDP can only achieve decent MOS scores when the target speakers speak their native language. 
On the contrary, the proposed model achieves the best MOS scores in all utterances except for the similarity score of CN utterances of speaker $d1$-CN-M. 
This suggests that the variance adaptors can further help the model disentangle speaker and language representations. It should be noted that these variance adaptors are originally proposed to ease the one-to-many mapping problem in non-autoregressive TTS models \cite{ren2021fastspeech}. Yet, our findings suggest that explicitly factorizing speech variations help disentangle language representations from other representations in cross-lingual TTS, even when we don't use any supervision or adversarial method on these hidden representations.

Finally, one can still observe a gap between the scores of target speakers speaking native and foreign languages. We speculate that, since the reference utterances are always in speakers' native language, the raters might be affected by the accent of the speaker or other bias factors when listening to samples.

\subsection{Ablation studies}
We perform ablation studies to (1) evaluate how adversarial training method would affect the proposed model, (2) verify the effectiveness of the two variance adaptors of the proposed model. For adversarial training, we add a gradient reversal layer (GRL) and a speaker classifier after encoder output \cite{zhang2019learning, shang21_interspeech}. We adopt the same speaker classifier and scale
factor, $\lambda$, for GRL as in \cite{zaidi2021daft}.

We report the MOS score of speech naturalness and speaker similarity on all kinds of utterances. The result is presented in Table \ref{tab:table3}. It can be seen that adversarial training has little effect on the proposed model. We hypothesize that IPA input and the diversity of speakers in the training dataset have already disentangled most language and speaker information. So the model can only obtain little information from the GRL layer. Furthermore, we empirically find that the GRL layer is sensitive to the scale factor, $\lambda$. Thus, the hyper-parameters we use may not be optimal for cross-lingual TTS and a more careful tuning might lead to a better result. Even so, the proposed model can achieve decent cross-lingual performance without any complex training method or hyper-parameter tuning.

On the other hand, a missing of the energy and pitch predictors leads to a significant drop in both naturalness and speaker similarity scores, which demonstrates the effectiveness of these variance adaptors in cross-lingual TTS.

\begin{table}[htb]
\centering
\caption{Overall naturalness and speaker similarity of ablation studies. Energy \& pitch denotes the energy predictor and pitch predictor, respectively.}
\begin{tabular}{p{12mm}p{12mm}p{12mm}p{12mm}p{12mm}}
\toprule
\multirow{2}{*}{Model} &   \multicolumn{2}{c}{$d1$-CN-M}  &   \multicolumn{2}{c}{$d1$-EN-F}         \\ \cline{2-5}  
& Naturalness  & Similarity  & Naturalness    & Similarity           \\ \midrule
\multirow{1}{*}{Proposed}  
                 & \textbf{4.19$\pm $0.24}  & \textbf{3.55$\pm $0.38}   
                 & 4.03$\pm $0.23    & 3.32$\pm $0.43      
                                \\ \hline
\multirow{1}{*}{+ GRL}  
                 & 4.19$\pm $0.24  & 3.54$\pm$0.38   
                 & \textbf{4.04$\pm $0.24}   & \textbf{3.33$\pm $0.44}     \\ 
\multirow{1}{*}{- energy}  
                 & 4.10$\pm $0.28  & 3.40$\pm$0.42   
                 & 4.02$\pm $0.23    & 3.29$\pm $0.44     \\ 
\multirow{1}{*}{- pitch}  
                 & 3.83$\pm $0.38  & 3.28$\pm$0.46   
                 & 3.54$\pm $0.33    & 3.06$\pm $0.44     \\ 

\bottomrule
\end{tabular}
\label{tab:table3}
\end{table}

\section{Conclusions}
\label{sec:conclude}

In this paper, we study the disentanglement of speaker and language representations in non-autoregressive cross-lingual TTS from various aspects. We propose the phoneme length regulator, which facilitates the implementation of non-autoregressive cross-lingual TTS models with IPA input representations using monolingual force aligners. We build a FastPitch-based model with IPA input that achieves decent speech naturalness and speaker similarity without any complex adversarial training method. Our experimental results show that an increasing number of training speakers, the IPA input representations and the variance adaptors in \cite{ren2021fastspeech} can help non-autoregressive cross-lingual TTS models disentangle speaker and language representations. 





In future work, we will investigate better methods to model the accent of target speakers in cross-lingual TTS, closing the gap between the native and foreign language of target speakers. We will also investigate better methods to model the prosody and emotions of speech in cross-lingual TTS.

\newpage

\bibliographystyle{IEEEtran}

\bibliography{mybib}


\end{document}